# Alternative Pathways for Multiple Exciton Generation Solar Cells by Tandem Configurations

**Jongwon Lee**

Independent Researcher, Seoul, South Korea
E-mail: jlee159@asu.edu

**Abstract:[1]** Multiple exciton generation solar cells (MEGSCs) undergo low efficiency due to material imperfections such as nonradiative recombination This paper introduces alternative approaches for realizing photovoltaic (PV) devices similar to MEGSCs. Furthermore, we reorganize the detailed balance equation of MEGSCs such that it is similar to that of independent connection tandem solar cells. This is possible because of the spectral dependence of the ideal QY. Finally, we compare these two similar equations and propose alternative approaches for realizing MEGSC-like tandem solar cells. We explain the difficulty in fabricating MEGSCs, which arises from the high rate of non-idealities. In this regard, the deconstruction of the detailed balance equation of MEGSCs can reveal alternative paths for replacing MEGSCs with tandem solar cell configurations.

## 1. Introduction

Multiple exciton generation solar cells (MEGSCs) are one of promising photovoltaic (PV) devices due to creating many electron and hole pairs (EHPs) per an incoming photon by impact ionization. Therefore, it can theoretically overcome the Shockley–Queisser limit due to the reduced thermalization losses [1,2,3]. After initial observation from Kolondinski et al.[2] in bulk Si solar cells (quantum efficiency of over 100%), its related researches have widely explored through QY extractions by pump-probe measurements in nanostructures solar cells (e.g., quantum dots) because of excellent quantum confinement [3,4]. Various materials have tested and its related QY demonstrates up to 300~700% (PbSe QD [5,6], CdTe [7], CdSe [8], InAs [9],[10], and Si [11]). However, the most results of QY extractions depends on surface state of materials [12-18] which it generates nonradiative recombination paths of the actual MEGSCscause. Thus, it has shown the low power conversion efficiencies [19-23].The development of near-perfect nanostructured materials is crucial because there exist non-idealities in MEGSCs [2,24-27]. Therefore, further development of nanostructured solar cells with enhanced material quality is necessary to ensure that negligible defects occur in MEGSCs. However, it might be difficult to develop such materials owing to the weak quantum confinement of carriers and related defects. Therefore, we propose an alternative approach for MEGSC fabrication based on the detailed balance (DB) limit of MEGSCs. The DB equation serves as an optimal testbed for the proposed solar cell structure with regard to predicting its performance. The ideal MEGSC can generate "m" EHPs under an energy variation from $m \cdot E_g$ to $(m+1) \cdot E_g$, where m is the IQY (m=1,2,3...) and $E_g$ is the bandgap energy; it is noteworthy that IQY exhibits a strong spectral dependence. The corresponding chemical potential is "$m \cdot q \cdot V$" in the recombination current term in the DB equation for MEGSCs, where q is the elementary charge and V is the operating voltage. Therefore, we can reconfigure the DB equation of MEGSCs such that it is similar to that of tandem solar cells, based on the spectral dependence of IQY. For instance, three EHPs are generated under the photon energy range of $3E_g$ to $4E_g$, and in this case, the corresponding chemical potential is $3 \cdot q \cdot V$. Thus, we can divide triple junction tandem solar

---

cell and its corresponding chemical potentials are qV, 2qV and 3qV. Based on spectral splitting, we can divide the solar spectrum from m·$E_g$ to (m+1)·$E_g$ and deconstruct the DB equation for MEGSCs such that it is similar to that for tandem solar cells under light concentration condition. Consequently, MEGSCs can be employed to form independent connected tandem solar cells (ICTSCs). Thus, we can replace the DB equation of MEGSCs with that of ICTSCs based on the same spectral splitting approach.

Notably, even if substantial experimental achievements are made regarding the QY, it will remain difficult to implement efficient MEGSCs without a near-perfect solar cell material. The deconstruction of the DB equation of MEGSCs provides alternative pathways, thereby serving as a means to overcome the aforenoted issue. Namely, it allows MEGSCs to be employed as tandem solar cells owing to the spectral dependence of IQY. Through the intensive analysis of the DB equations corresponding to these two cell types (MEGSC and tandem solar cells), we can determine their similarities and propose practical methods for fabricating effective solar cells. Notably, tandem solar cells have experienced technological advancements in terms of their growth, and interface defects have arisen owing to multi-stacked growth, which is performed to improve tandem cell efficiency [28,29,30]. These limitations hinder further developments; in this regard, spectral splitting using holographic optics can serve as a good replacement for vertical stacked tandem solar cells to mitigate lattice mismatch, diversify material selection, and increase the number of junctions [28,29,30]. This will expand the configuration options for tandem solar cells. Thus, one can employ the optical spectrum splitting scheme to realize novel tandem solar cells [28,29,30]. Thus, we propose novel tandem solar cell structures based on the reconfiguration of the DB of MEGSCs.

## 2. Theory

In this section, we will discuss the DB equations of MEGSCs to determine other possible methods of fabricating MEGSCs; next, we describe the procedure of converting MEGSCs to ICTSCs. First, from Eq. (1) to Eq. (4), we present the DB equation of MEGSCs. The IQY has a spectral dependence in that an MEGSC can generate "m" EHPs from photons with an energy ranging from m·$E_g$ to (m+1)·$E_g$. In addition, the corresponding chemical potential is m·q·V. Based on the integrand of the recombination current in MEGSCs, an exponential term, namely, exp[m·($E_g$-q·V)], is obtained. Thus, the effective chemical potential is "q·V" instead of "m·q·V." To avoid mathematical errors with regard to the recombination current, the open circuit voltage ($V_{OC}$) must be smaller than <$E_g$ [31].

$$\text{Ideal QY(E)} = \begin{cases} 0 & 0 < E < E_g \\ m & m \cdot E_g < E < (m+1) \cdot E_g \quad m = 1,2,3 \dots \\ M & E \geq M \cdot E_g \end{cases} \tag{1}$$

where m is the number of multiple EHPs generated, M is the maximum number of EHPs, $E_g$ is the bandgap, and E is the photon energy.

$$\phi(E_1, E_2, T, \mu) = \frac{2\pi}{h^3 c^2} \int_{E_1}^{E_2} \frac{E^2}{\exp[(E-\mu)/kT]-1} dE \tag{2}$$

$$\phi_{MEG}(E_1, E_2, T, \mu) = \frac{2\pi}{h^3 c^2} \int_{E_1}^{E_2} \frac{QY(E) \cdot E^2}{\exp[(E-\mu_{MEG})/kT]-1} dE \tag{3}$$

$$J_{MEG} = \sum_{m=1}^{M-1} \begin{bmatrix} q \cdot C \cdot f_s \cdot \phi_{MEG}(m \cdot E_g, (m+1) \cdot E_g, T_S, 0) \\ +q \cdot C \cdot f_s \cdot \phi_{MEG}(m \cdot E_g, (m+1) \cdot E_g, T_S, 0) \\ -q \cdot \phi_{MEG}(m \cdot E_g, (m+1) \cdot E_g, T_C, q \cdot m \cdot V) \end{bmatrix} + \begin{bmatrix} q \cdot C \cdot f_s \cdot \phi_{MEG}(M \cdot E_g, \infty, T_S, 0) \\ +q \cdot C \cdot (1-f_s) \cdot \phi_{MEG}(M \cdot E_g, \infty, T_S, 0) \\ -q \cdot \phi_{MEG}(M \cdot E_g, \infty, T_C, q \cdot M \cdot V)) \end{bmatrix} \tag{4}$$

where $\phi$ is the particle flux given by Planck's equation for a temperature T with a chemical potential $\mu$ in the photon energy range of $E_1$ to $E_2$. Furthermore, h is Planck's constant, c is the speed of light, and $\mu$ is the chemical potential (=q·V), where V is the operating voltage. Moreover, $\mu_{MEG}$ is the chemical potential of MEG (=q·QY(E)·V), k is the Boltzmann constant, J is the current density of the solar cell, q is the charge of an elementary particle, C is the optical concentration, $f_s$ is a geometry factor (1/46200), $T_S$ is the temperature of the sun (6000 K), and $T_C$ is the temperature of the solar cell (300 K).

Based on Eq. (4), we can reorganize the DB equation of MEGSCs as Eq. (5) and (6) by changing the independent chemical potential, $q \cdot m \cdot V_{(m),RMT}$. That is, we can consider that this equation corresponds to the reconfigured MEG for tandem (RMT) approaches, except that the actual bandgap energy in the latter is $E_g$ instead of $m \cdot E_g$.

$$\phi_{RMT}(E_1, E_2, T, \mu) = \frac{2\pi}{h^3 c^2} \int_{E_1}^{E_2} \frac{m \cdot E^2}{\exp[(E - q \cdot m \cdot V_{(m),RMT})/kT] - 1} dE \tag{5}$$

$$J_{RMT(M)} = \begin{bmatrix} Q \cdot C \cdot F_S \cdot \Phi_{RMT}(M \cdot E_G, (M+1) \cdot E_G, T_S, 0) \\ + Q \cdot C \cdot F_S \cdot \Phi_{RMT}(M \cdot E_G, (M+1) \cdot E_G, T_S, 0) \\ - Q \cdot \Phi_{RMT}(M \cdot E_G, (M+1) \cdot E_G, T_C, Q \cdot M \cdot V_{(M),RMT}) \end{bmatrix} \tag{6}$$

Based on spectral splitting, we can divide the spectrum from $m \cdot E_g$ to $(m+1) \cdot E_g$, where $m=1,2,3\ldots M-1$, M (here, M corresponds to the top-most junction). Thus, we can reconfigure MEGSCs as tandem solar cells (see Eq. (6) ) with "M" junctions after the deconstruction of the DB equation of MEGSCs. Based on the RMT approach, "m" represents the number corresponding to a junction with the fundamental bandgap, $E_g$. Thus, we can define the incident photon energy as $m \cdot E_g$. Furthermore, the chemical potential corresponding to each junction, $m \cdot q \cdot V$, can be elucidated by the fact that the fundamental chemical potential, $q \cdot V$, is multiplied by "m." To distinguish the chemical potential for MEGSCs, we define $\mu_{(m),RMT} = q \cdot m \cdot V_{(m),RMT} < m \cdot E_g$. Correspondingly, $V_{OC(m),RMT} < E_g$. Therefore, $V_{(m),RMT}$ is the optimum chemical potential of each cell without regarding "m." Furthermore, its short circuit current density will increase by "m" based on Eq. (2), and the chemical potential is $qV_{(m),RMT}$. Fig. 3 shows the procedure of the conversion of the DB equations of MEGSCs to those of ICTSCs. Therefore, Eq. (7) describes multiple single junction solar cells with the same bandgaps $(=E_g)$ and varied incident photon energies, $m \cdot E_g$ corresponding to each sub-cell. If "m" is removed or neglected, except for the chemical potential in Eq. (7), we could obtain the DB equations of independent connected tandem solar cells (ICTSC1), with $\mu_{(m),ICTSC1} = q \cdot m \cdot V_{(m),RMT}$. With this consideration, we can set Eq. (8) for ICTSC1. Note that the $V_{OC}$ for ICTSC1 is less than $m \cdot E_g$.

In summary, we establish three models to generate the tandem configurations of MEGSCs.

(a) Reconfigured MEGSCs to tandem (RMT): from the DB limit of MEGSCs, we can sort or group light-generated and recombination current from spectrum splitting from $m \cdot E_g$ to $(m+1) \cdot E_g$

(b) ICTSC1: After removal of "m," except for the chemical potential in RMT, the entire equation could be Eq. (18), and it is similar to conventional independent connected tandem solar cells.

(c) ICTSC2: Conventional independent connected tandem solar cells

$$J_{RMT} = \begin{bmatrix} q \cdot C \cdot f_s \cdot \phi_{RMT}(E_g, 2 \cdot E_g, T_S, 0) \\ + q \cdot C \cdot (1 - f_s) \cdot \phi_{RMT}(E_g, 2 \cdot E_g, T_S, 0) \\ - q \cdot \phi_{RMT}(E_g, 2 \cdot E_g, T_C, \mu_{(1),RMT}) \end{bmatrix} + \cdots + \begin{bmatrix} q \cdot C \cdot f_s \cdot \phi_{RMT}(M \cdot E_g, \infty, T_S, 0) \\ + q \cdot C \cdot (1 - f_s) \cdot \phi_{RMT}(M \cdot E_g, \infty, T_S, 0) \\ - q \cdot \phi_{RMT}(M \cdot E_g, \infty, T_C, \mu_{(M),RMT}) \end{bmatrix} \tag{7}$$

where $\mu_{(m),RMT} = q \cdot m \cdot V_{RMT(m)}$ is the chemical potential corresponding to the RMT approach, $V_{(m),RMT(m)} < E_g$.

$$J_{ICTSC1} = \begin{bmatrix} q \cdot C \cdot f_s \cdot \phi(E_g, 2 \cdot E_g, T_S, 0) \\ + q \cdot C \cdot (1 - f_s) \cdot \phi(E_g, 2 \cdot E_g, T_S, 0) \\ - q \cdot \phi(E_g, 2 \cdot E_g, T_C, \mu_{(1),RMT}) \end{bmatrix} + \cdots + \begin{bmatrix} q \cdot C \cdot f_s \cdot \phi(M \cdot E_g, \infty, T_S, 0) \\ + q \cdot C \cdot (1 - f_s) \cdot \phi(M \cdot E_g, \infty, T_S, 0) \\ - q \cdot \phi(M \cdot E_g, \infty, T_C, \mu_{(M),RMT}) \end{bmatrix} \tag{8}$$

Eq. (9) is the conventional DB equation of ICTSC2. Note that the chemical potential terms in Eq. (9) and Eq.(8) can be nearly identical. This will be described further in the Results section herein.

$$J_{ICTSC2} = \begin{bmatrix} q \cdot C \cdot f_s \cdot \phi(E_g, 2 \cdot E_g, T_S, 0) \\ + q \cdot C \cdot (1 - f_s) \cdot \phi(E_g, 2 \cdot E_g, T_S, 0) \\ - q \cdot \phi(E_g, 2 \cdot E_g, T_C, \mu_{(1),ICTSC2}) \end{bmatrix} + \cdots + \begin{bmatrix} q \cdot C \cdot f_s \cdot \phi(M \cdot E_g, \infty, T_S, 0) \\ + q \cdot C \cdot (1 - f_s) \cdot \phi(M \cdot E_g, \infty, T_S, 0) \\ - q \cdot \phi(M \cdot E_g, \infty, T_C, \mu_{(M),ICTSC2}) \end{bmatrix} \tag{10}$$

where $\mu_{(m),\text{ICTSC}} = q \cdot V_{(m)}$ is the chemical potential corresponding to the RMT approach.

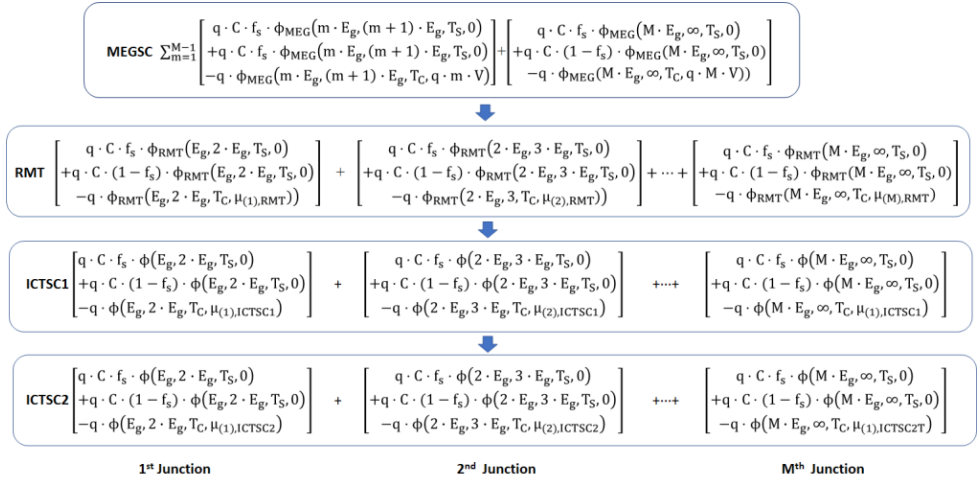

Fig. 1 Procedure of conversion of the DB equation from that of a MEGSC to those of ICTSC1 and ICTSC2, where m=1,2,3,…M-1,M (maximum QY). By employing the spectral splitting approach, one can deconstruct each spectral group corresponding to the MEGSC, and the counterpart for the RMT approach can be obtained. Next, we remove "m," except for the chemical potential, $\mu_{(m),\text{RMT}} = q \cdot m \cdot V_{(m),\text{RMT}}$ to obtain ICTSC1. The chemical potential term changes to $\mu_{(m),\text{ICTSC}1} = q \cdot m \cdot V_{(m),\text{RMT}}$. For a conventional ICTSC, the chemical potential is defined as $\mu_{(m),\text{ICTSC}2}$.

Figure 1 describes the procedure of conversion of the DB equation of a MEGSC to that of an ICTSC.

(1) The DB equation of a MEGSC can be deconstructed based on spectral splitting corresponding to different chemical potentials.

(2) After removing the quantum yield except for the chemical potential, this equation becomes that corresponding to ICTSC1, with $\mu_{(m),\text{ICTSC}1} = q \cdot m \cdot V_{(m),\text{RMT}} < m \cdot E_g$ for each junction.

In Table 1, RMT, ICTSC1, and ICTSC2 are compared to verify the proposed approach and summarize the chemical potentials corresponding to the three configurations. The chemical potentials for ICTSC1 and ICTSC2 are $q \cdot V_{(m),\text{RMT}}$ and $q \cdot V_{(m),\text{ICTSC}2}$, respectively; these must be less than $m \cdot E_g$ as this represents the corresponding bandgap energy of each junction, except for RMT.

Table 1. Comparison of chemical potential and $V_{OC}$ and the corresponding bandgap energy of each junction for RMT, ICTSC1, and ICTSC2.

| | Short Circuit Current | Chemical Potential | $V_{OC}$ | Bandgap Energy of each junction |
|---|---|---|---|---|
| RMT | $q \cdot C \cdot f_S \cdot \phi_{\text{RMT}}(mE_g,(m+1)E_g,T_S,0)$ $+ q \cdot C \cdot (1-f_S) \cdot \phi_{\text{RMT}}(mE_g,(m+1)E_g,T_C,0)$ | $\mu_{(m),\text{RMT}} = q \cdot V_{(m),\text{RMT}} < E_g$ | $V_{OC(m),\text{RMT}}$ $< E_g$ | $E_g$ |
| ICTSC1 | $q \cdot C \cdot f_S \cdot \phi(mE_g,(m+1)E_g,T_S,0)$ $+ q \cdot C \cdot (1-f_S) \cdot \phi(mE_g,(m+1)E_g,T_C,0)$ | $\mu_{(m),\text{ICTSC}1} = q \cdot m \cdot V_{(m),\text{RMT}} < m \cdot E_g$ | $V_{OC(m),\text{ICTSC}1}$ $< m \cdot E_g$ | $m \cdot E_g$ |
| ICTSC2 | $q \cdot C \cdot f_S \cdot \phi(mE_g,(m+1)E_g,T_S,0)$ $+ q \cdot C \cdot (1-f_S) \cdot \phi(mE_g,(m+1)E_g,T_C,0)$ | $\mu_{(m),\text{ICTSC}2} = q \cdot V_{(m),\text{ICTSC}2} < m \cdot E_g$ | $V_{OC(m),\text{ICTSC}2} < m \cdot E_g$ | $m \cdot E_g$ |

## 3. Results

In this section, we present and compare the limiting efficiencies corresponding to ICTSC1, ICTSC2, and RMT. After analyzing the DB limit of MEGSCs, the spectral dependence (or spectral division) could be employed to reconfigure MEGSCs as tandem solar cells. With regard to the calculations related to tandem solar cells, each junction will emit its photon energy; therefore, the topmost junction (M$^{th}$ junction) will emit the photon energy of M·E$_g$. Considering ideal MEG, after Auger recombination, the last electron will emit a photon energy of M·E$_g$ (via radiative recombination) with a chemical potential of q·M·V without carrier losses. The RMT approach entails multiple single junction cells that absorb photons with varied photon energies, m·E$_g$; in this regard, this approach is similar to that of tandem solar cells. However, this approach is only available under the specific conditions described herein. After reconfiguring RMT to ICTSC1, the chemical potentials of each junction, q·m·V$_{(m),RMT}$, including that of the last junction (M$^{th}$ junction) will lead to the emission of photon energy of M·E$_g$. The maximum efficiency and optimum E$_g$ of MEGSCs are 85.9% and 0.05 eV, respectively, with a maximum QY of 200 under full-light concentration. Thus, we set 200 junctions (m=1,2,3…,199,200) with E$_g$ = 0.05 eV via the ICTSC1, ICTSC2, and RMT approaches, and the corresponding results are shown in Fig. 5 and Table. 2. In RMT approach, first, we neglect the "m" term in the chemical potential (see Fig, 2(a)), such that the chemical potential is below the bandgap energy of first junction, E$_g$. This result is based on μ$_{(m),RMT}$=q·m·V$_{(m),RMT}$ < m·E$_g$. Thus, after removing m, q·V$_{(m),RMT}$ becomes less than E$_g$. In ICTSC1, the chemical potentials obtained by multiplying "m" correspond to the chemical potentials at the junctions corresponding to m·E$_g$, and they are nearly identical to those obtained via the ICTSC2 approach (see Fig. 2 (b)). Moreover, the maximum powers obtained via the three approaches (ICTSC1, ICTSC2, and RMT) are approximately the same (see Fig. 2(c)). The corresponding calculated efficiencies are approximately 86%, and this value approaches that in the case of an infinite number of junctions. Typically, the increased short circuit current of RMT leads to results that are similar to those obtained for ICTSC1 and ICTSC2, even if there are multiple single junction (same bandgap energy) solar cells operating under different incoming photon energies.

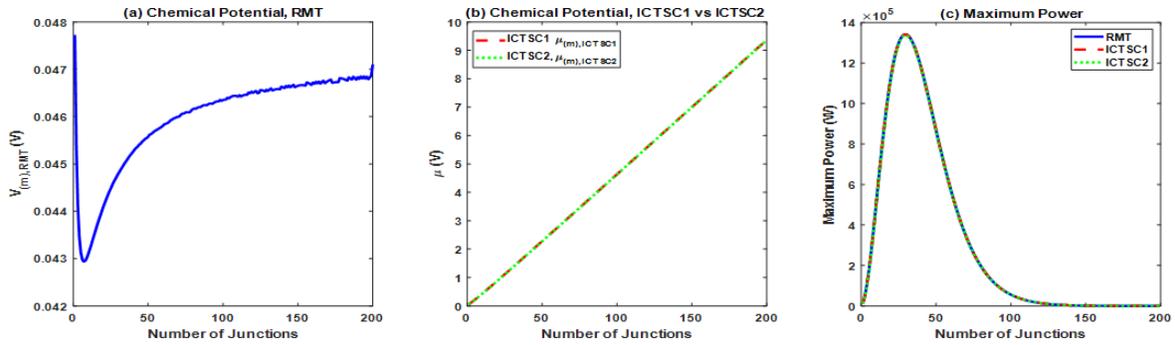

Fig. 2 Comparison of ICTSC1, ICTSC2, and RMT approaches for 200 junctions; here, E$_g$=0.05 eV under full-light concentration. Furthermore, the corresponding bandgaps are given by m·E$_g$, where m is the number of junctions (=1,2,3…199, 200).

Table. 2 Simulation results for ICTSC1, ICTSC2, and RMT for 200 junctions and MEGSC for 200 EHPs; here, E$_g$=0.05 eV under the full-light concentration.

| Approach | ICTSC1 | ICTSC2 | RMT | MEG |
|---|---|---|---|---|
| Efficiency (%) | 86.3% | 86.5% | 86.5% | 85.9% |

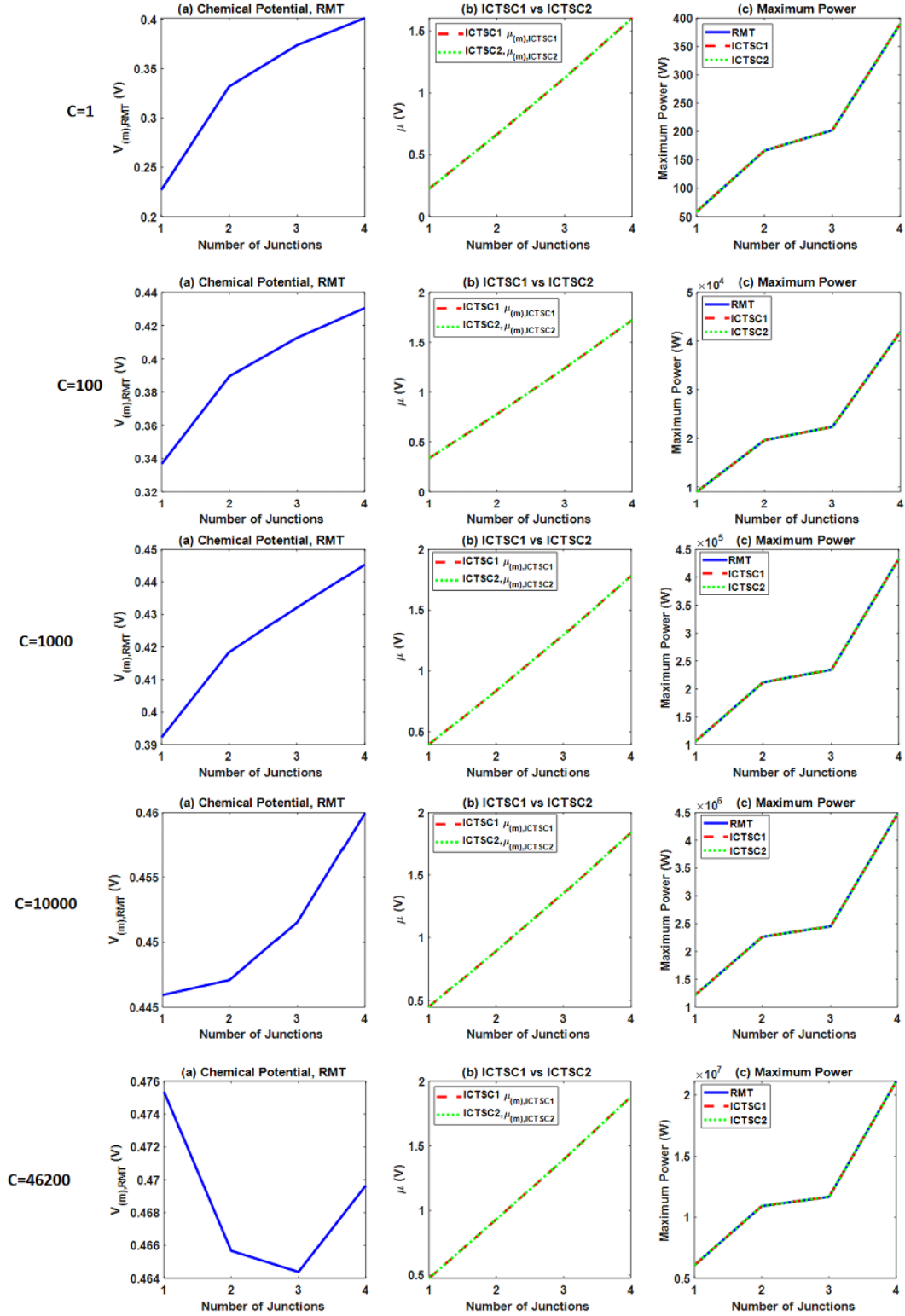

Fig. 3 Comparison between ICTSC1, ICTSC2, and RMT approaches in the case of quadruple junctions; here, $E_g$=0.5 eV under C=1,100, 1000, 10000 add 46200 suns. Furthermore, the corresponding bandgaps are given by m·$E_g$, where m is the number of junctions (=1,2,3,4).

The aforenoted results are difficult to achieve in a practical solar cell owing to the required number of junctions being excessive. Thus, we employ a small number of junctions to compare both approaches. In this case, we consider quadruple junctions where the first junction has $E_g$= 0.5 eV; thus, the bandgaps of the remaining junctions are 1 eV, 1.5 eV, and 2 eV. The chemical potential of each junction and the corresponding maximum power are also nearly identical under the full-light concentration (See Fig. 3 and Table 3). Furthermore, the corresponding average efficiency is approximately 67.7%, which is nearly the same as that obtained in the case of MEGSCs. And, we provide additional conditions of sun-light (ex, C=1, 100, 1000, 10000 and 46200) to compare MEGSC. While comparing RMT approaches and ICTSC, there is no dependence of light-concentration that all the results are nearly identical. However, after comparing MEGSC, we could find a strong light-concentration dependence from these cases. Under one sun and C=100, these two results show difference of more than 5% of theoretical efficiencies between ICTSC and MEGSC. However, after increasing the light-concentration, its difference become merged to nearly 0 that this ICTSC approach can be valid under the moderate and high light-concentration condition (see Table 3).

Table. 3 Simulation results obtained for ICTSC1, ICTSC2, RMT, and MEGSC in the case of quadruple junctions where $E_g$=0.5 eV under the light concentration (1, 100, 1000, 10000, 46200) where $\eta$ is efficiency (%)

| Concentration | ICTSC1($\eta$) | ICTSC2($\eta$) | RMT($\eta$) | MEG($\eta$) |
|---|---|---|---|---|
| 1 | 51.35 | 51.35 | 51.36 | 37.36 |
| 100 | 58.38 | 58.38 | 58.39 | 53.38 |
| 1000 | 61.92 | 61.92 | 61.93 | 60.22 |
| 10000 | 65.44 | 65.44 | 65.45 | 65.20 |
| 46200 | 67.71 | 67.71 | 67.73 | 67.65 |

These results demonstrate that the DB model obtained via the RMT approach can replace that of the ICTSCs under the moderate and high-light concentration condition. Furthermore, the performances obtained via this approach were similar to those of MEGSCs under mid or high light concentration condition. That is, ICTSCs can exhibit performance that is nearly identical to that of MEGSCs under the full-light concentration; this is evidenced by the fact that both exhibited similar maximum powers.

Based on these results, we suggest the following methods for fabricating ICTSC PV devices similar to MEGSCs. (1) We can develop mechanically stacked tandem solar cells (i.e., independent connected tandem solar cells) with bandgaps of m·$E_g$ (see Fig .4 (a)). (2) We can fabricate multiple islands having a bandgap energy of m·$E_g$ on a single substrate (ex, see Fig 4 (b), that is, a lateral array of multiple-bandgap solar cells [29,30]). (3) Materials with different bandgaps can be used by selectively matching their m·$E_g$ values to obtain a single configuration based on option (2). Typically, to realize PV cells based on method (2), nanostructures with various sizes can be employed, as the bandgap energy is tunable. For instance, if $E_g$ is 1.1 eV, which is the case for Si, we can fabricate materials with two bandgaps (1.1 eV, 2.2 eV) by adjusting the size of the nanostructure. Using combinations of these two materials, we can fabricate a two-junction mechanically stacked tandem solar cell or two isolated islands on a Si substrate. The latter case may require specific optical equipment for spectrum splitting to enable the absorption of a particular wavelength of light so that the system can be implemented based on the ICTSC approach. Fig. 4 depicts the proposed solar cell structures.

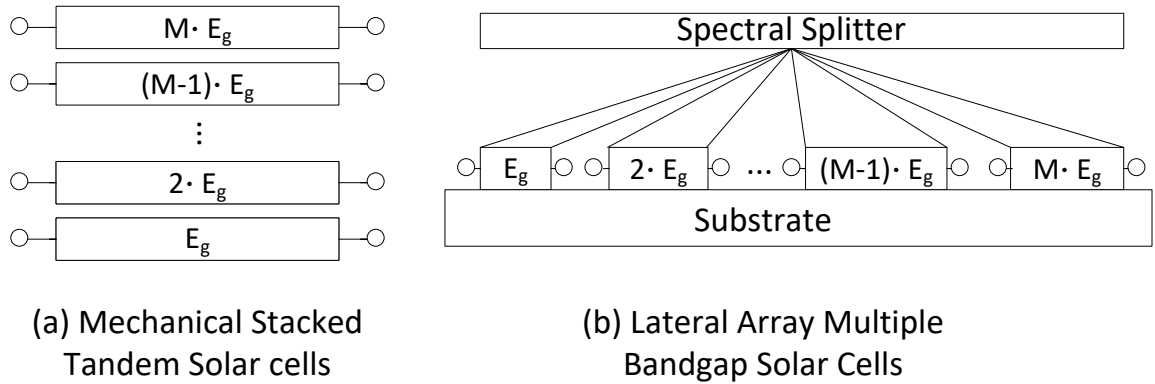

(a) Mechanical Stacked Tandem Solar cells

(b) Lateral Array Multiple Bandgap Solar Cells

Fig. 4 Schematics of two types of tandem solar cells for achieving ICTSC-like MEGSCs. For instance, if $E_g$=1.12 eV, we can construct two ($E_g$, $2E_g$) or three junctions ($E_g$, $2E_g$, and $3E_g$) using mechanically stacked tandem solar cells or lateral array multiple bandgap solar cells.

## 4. Conclusions

Notably, it is difficult to realize effective MEGSCs owing to the aforenoted imperfections, which lead to non-radiative recombination or non-idealities. Thus, we proposed a tandem approach based on the spectral dependence of the IQY of MEGSCs. Based on the DB equation of MEGSCs and the spectral dependence of IQY, the proposed method can be applied to tandem solar cells via spectral splitting approaches. Using the proposed approach, we deconstructed the DB equation of MEGSCs such that it was similar to that of ICTSCs and compared these concepts. The maximum powers corresponding to each junction are nearly identical under light concentration. Furthermore, it was found that the maximum efficiencies obtained via the ICTSC1, ICTSC2, and RMT approaches are also nearly the same as those obtained for MEGSCs under light concentration conditions. Thus, the ICTSC approach can replace MEGSCs under the specified light concentration condition. Finally, we proposed three approaches for realizing ICTSCs similar to MEGSCs, namely, (1) the use of a mechanically stack structure and (2) multiple isolated islands on a single substrate with optical spectrum splitting. Note that even if ideal conditions are assumed in the DB equations of RMT, tandem solar cells, and MEGSCs, the RMT and ICTSC equations can be employed to propose the next steps for producing MEGSC-like tandem solar cells. In summary, through this research, practical approaches for developing MEGSC-like tandem solar cells were proposed. We believe that the results of the present study can be used to develop novel, effective solar cells.

**Author Contributions:** Dr. Jongwon Lee suggested the main idea, wrote, and reviewed this article.

**Funding: N/A**

**Acknowledgments:** N/A